# Epitaxially defined Luttinger liquids on MoS$_2$ bicrystals

Bingchen Deng[1,2], Heonsu Ahn[3,4], Jue Wang[1,2], Gunho Moon[3,4], Ninad Dongre[5], Chao Lei[6], Giovanni Scuri[1,2,7], Jiho Sung[1,2], Elise Brutschea[1], Kenji Watanabe[8], Takashi Taniguchi[9], Fan Zhang[5†], Moon-Ho Jo[3,4,10†], and Hongkun Park[1,2†]

[1]Department of Chemistry and Chemical Biology, Harvard University, Cambridge, MA 02138, USA

[2]Department of Physics, Harvard University, Cambridge, MA 02138, USA

[3]Center for Van der Waals Quantum Solids, Institute for Basic Science (IBS), Pohang 37673, Korea

[4]Department of Materials Science and Engineering, Pohang University of Science and Technology (POSTECH), Pohang 37673, Korea

[5]Department of Physics, The University of Texas at Dallas, Richardson, TX 75080, USA

[6]Department of Physics, The University of Texas at Austin, Austin, TX 78712, USA

[7]Department of Electrical Engineering, Stanford University, Stanford, CA 94305, USA

[8]Research Center for Electronic and Optical Materials, National Institute for Materials Science, 1-1 Namiki, Tsukuba 305-0044, Japan

[9]Research Center for Materials Nanoarchitectonics, National Institute for Materials Science, 1-1 Namiki, Tsukuba 305-0044, Japan

[10]Department of Physics, Pohang University of Science and Technology (POSTECH), Pohang 37673, Korea

†To whom correspondence should be addressed: zhang@utdallas.edu, mhjo@postech.ac.kr, and hongkun_park@harvard.edu

**ABSTRACT**

A mirror twin boundary (MTB) in a transition metal dichalcogenide (TMD) monolayer can host one-dimensional electron liquid of a topological nature with tunable interactions. Unfortunately, the electrical characterization of such boundaries has been challenging due to the paucity of samples with large enough size and high quality. Here, we report an epitaxial growth of monolayer molybdenum disulfide ($MoS_2$) bicrystals with well-isolated MTBs that are tens of micrometers long. Conductance measurements of these MTBs exhibit power-law behaviors as a function of temperature and bias voltage up to room temperature, consistent with electrons tunneling into a Luttinger liquid. Transport measurements of two distinct types of MTBs reveal the critical role of the atomic-scale defects. This study demonstrates that MTBs in TMD monolayers provide an exciting new platform for studying the interplay between electronic interactions and topology.

A mirror twin boundary (MTB) in a transition metal dichalcogenide (TMD) monolayer is a one-dimensional (1D) grain boundary that separates two intersecting domains with 180° in-plane rotation [1-3]. Density functional theory calculations predict that MTBs in a TMD monolayer should host 1D electron liquids [3] of topological origin and thus provide an interesting system for investigating the interplay between electronic correlation and topology [4,5]. Although scanning tunneling microscopy studies of a short (~10 nm) MTB segment [6-10] and an angle-resolved photoemission spectroscopy study of a dense MTB network [11] have provided a glimpse of correlated electron states on these 1D boundaries, sample limitations have posed an outstanding challenge for electron transport experiments on MTBs.

In this work, we epitaxially grow monolayer molybdenum disulfide ($MoS_2$) bicrystals on a c-plane sapphire substrate. These bicrystals possess MTBs of tens of micrometers in length at the intersections of two $MoS_2$ monolayer single crystals. We identify two types of MTBs: Type I, in which the MTB is nearly straight along the whole intersection of the two crystals, and Type II, in which the MTB makes frequent 60° turns, producing a zigzag structure. Field-effect transistor devices based on Type I and Type II MTBs exhibit distinct conductance behaviors: Type I MTBs exhibit Coulomb blockade at cryogenic temperatures, indicative of fewer defects along the 1D channel, while Type II MTBs do not. Both Type I and Type II MTBs show power-law behaviors of tunneling conductance as a function of temperature and bias voltage up to room temperature, consistent with electrons tunneling into Luttinger liquids. The transport behaviors of Type I and Type II MTBs are characterized by distinct power law exponents, however, pointing to the role of sharp turns (and the associated defects) in producing large momentum scattering and valley pseudospin mixing for Type II MTBs.

**Epitaxial growth of MoS$_2$ bicrystals**

When MoS$_2$ monolayers with a three-fold in-plane rotational symmetry are epitaxially grown on a c-plane sapphire with a six-fold rotational symmetry, two types of crystallographic variants can be formed. These two variants are exactly 180° rotated from each other and can establish the MTBs when they coalesce to form MoS$_2$ bicrystals during the growth. In our S-rich growth condition, each variant has triangular facets with S-terminating edges [marked with either blue or red triangle in Fig.1(a-b)]. As such, when they form a bicrystal, MTBs are essentially the type with a 4|4E sublattice, where each S chain along the grain boundary is a reflection copy of the other with a half-unit cell shift [Fig. 1(c)].

We achieved large-area bicrystal epitaxy using metal-organic chemical vapor deposition (MOCVD) with kinetics control [12] (see Supplemental Material [13] for details). Figure 1(a) is an optical image of such bicrystals, where the macroscopic shapes of the bicrystals can be classified as either rhombus or bowtie. In the rhombus bicrystals, where two triangular facets meet at the edges, a straight MTB (Type I) is formed, which extends over tens of micrometers in length. In the bowtie type, where two facets meet at the vertices, a zigzag MTB (Type II) is formed. Figure 1(b) and 1(c) show dark-field transmission electron microscopy (DF-TEM) image and high-angle annular dark-field scanning transmission electron microscopy (HAADF-STEM) images of these MTB types (see Supplemental Material [13] for details). In the zigzag MTBs, the straight 4|4E MTB segments (average length is ~40 nm) are connected by 60° turns. In each turn, we typically find one or two point defects, identified as 5|7 or 4|8 rings.

**Transport characterizations of Type I (straight) and Type II (zigzag) MTBs**

The as-grown $MoS_2$ flakes adhere strongly to the sapphire substrate, making them difficult to integrate into van der Waals heterostructures. To preserve their intrinsic properties via hexagonal boron nitride (hBN) encapsulation, we first transfer the $MoS_2$ flakes out of the sapphire by spin coating a polymethyl methacrylate (PMMA) layer on top and peeling the flakes off using a thermal release tape. We then transfer the tape to a $SiO_2$/Si chip and raise the temperature to release the flakes. After cleaning the chip with acetone to remove the PMMA, we use conventional dry transfer method [20] to encapsulate a monolayer $MoS_2$ bicrystal with top and bottom hBN layers. We add a graphite layer as electrostatic gate and lithographically define gold/titanium electrodes to complete field-effect transistor [Fig. 2(a), and see Supplemental Material [13] for detailed fabrication procedures].

The dashed lines in Fig. 2(b) illustrate the room-temperature drain-source current $I$ as a function of gate voltage ($V_G$) obtained from a control device that does not contain an MTB. It exhibits a regular *n*-type transistor behavior [21], as expected for a single-crystal $MoS_2$ monolayer device. The device **D1** containing a 200-nm-long Type I MTB shows a markedly different behavior. The current in **D1** extends deep into the negative $V_G$ region and remains only weakly dependent on $V_G$, signaling the MTB's metallic character. A 220-nm long Type II MTB transistor (**D2**) also exhibits measurable currents at negative $V_G$, but the current is strongly dependent on $V_G$, eventually turning off at very negative gate voltages [Fig. 2(c)]. In addition, the magnitude of the current in **D2** is generally much smaller than that of **D1**. All the transistors with the hBN passivation show negligible hysteresis with $V_G$. Details about the transport measurements are discussed in Supplemental Material [13].

When the temperature falls below $T = 70$ K, the current of **D1** exhibits dramatic oscillations as a function of $V_G$. Figure 2(d) shows the differential conductance ($dI/dV$) map as a function of both $V_G$ and bias voltage $V$ measured from **D1** at 5.0 K. There is a clear modulation of the conductance gap (purple region) as a function of $V_G$ at cryogenic temperatures, the hallmark of Coulomb blockade. This phenomenon is a consequence of the finite energy required to add (remove) a single electron to (from) the finite-sized Type I MTB [22]. The aperiodic pattern in Fig. 2(d) suggests there are multiple weakly coupled electron islands in series in this device [22-24]. An order-of-magnitude estimate of the geometrical capacitance gives a charging energy of ~26 meV for the 200-nm-long MTB [25-27]. The overall addition energies extracted from Fig. 2(d) are much larger than this estimate, consistent with the fact that the device consists of smaller electron islands in series. The formation of multiple electron islands may be due to occasional defects along the Type I MTB, e.g., a sequence of two 60° turns as shown in Supplemental Material Fig. S1 [13].

As shown in Fig. 2(e), **D2** does not exhibit current oscillations as a function of $V_G$ at 5.0 K, but instead the conductance gap gradually increases as $V_G$ becomes more negative. The absence of conductance gap modulation is consistent with the more defective nature of Type II MTB resulting from the frequent quasi-periodic 60° turns. In this case, the metallic character of MTB is compromised by these atomic-scale defects, and the conduction is eventually turned off.

As the temperature increases from 5.0 K to room temperature, the zero-bias conductance of **D1** gradually increases with temperature, and the device behavior changes from Coulomb blockade to a power-law type (linear in the log-log plot) above $T \sim 70$ K [Fig. 3(a)]. Such a change-over is reminiscent of the transport characteristics in carbon nanotube field-effect transistors changing from Coulomb blockade to Luttinger liquid physics [28,29]. Out of the blockade regime, we extract a power-law exponent of 0.9 (guided by the dashed line) over the temperature range of 70 to 297 K. In contrast, the zero-bias conductance of **D2** follows a power-law behavior over the whole measured temperature range with an exponent of 2.1 [Fig. 3(b)].

The power-law behaviors of conductance with temperature strongly suggest the formation of Luttinger liquids (correlated electronic states in 1D) on the MTBs. In a Luttinger liquid with tunnel contacts to metal electrodes, the tunneling density of states is suppressed following a power-law relationship with energy [22]. In our transport experiment, the energy comes from both temperature ($k_BT$) and bias voltage ($eV$). As a result, when the bias becomes dominant ($eV/k_BT \gg 1$), the differential conductance should also exhibit a power law behavior as a function of $V$, with the same power-law exponent as with temperature. In fact, the tunnel differential conductance should follow

$$\frac{dI}{dV} = AT^\alpha \cosh\left(\gamma \frac{eV}{2k_BT}\right) \left|\Gamma\left(\frac{1+\alpha}{2} + \gamma \frac{ieV}{2\pi k_BT}\right)\right|^2, \quad (1)$$

where $\alpha$ is the exponent and with $A$ and $\gamma$ as fitting parameters [28,30]. This equation suggests that in a scaled differential conductance-scaled bias plot ($dI/dV/T^\alpha$ vs. $eV/k_BT$), data collected at different temperatures and different biases should collapse to a single curve.

Figure 3(c) presents the $dI/dV/T^\alpha$ vs. $eV/k_BT$ plot for temperatures above 70 K (the Coulomb blockade-Luttinger liquid crossover) for **D1**. In the high-bias regime ($eV/k_BT \gg 1$), the differential conductance shows the expected power law exponent of 0.9 with bias voltage. Additionally, the data from all measured temperatures collapse to a single curve, consistent with the universal scaling law in Eq. 1. Using the exponent $\alpha$, it is possible to infer the electron-electron interaction strength, i.e., the Luttinger parameter $g$. For a non-interacting system, $g = 1$, and for a strongly correlated state with repulsive interactions, $0 < g < 1$. In our MTB devices with valley-chirality locking (see Fig. 4 below), because electrons tunnel into and out of the Fermi-liquid metal electrodes, we reach $\alpha = (g^{-1} + g - 2)/4$ (see Supplemental Material [13] for details). The exponent of 0.9 corresponds to $g = 0.185$, indicating strong electronic correlations in our MTB.

In Eq. 1, $\gamma$ is the ratio of the voltage drop across the most resistive tunneling junction to the total bias voltage $V$. Consequently, $1/\gamma$ gives the minimum number of junctions in the device. As shown in Fig. 3(c), in the best fitting of our observations to Eq. 1, we find $\gamma$ to be ~0.3: with two tunneling junctions to the metal electrodes, this suggests that the MTB in **D1** consists of approximately three Luttinger liquid segments and that there are tunneling events between them. This is consistent with the aperiodic Coulomb blockade [Fig. 2(d)], which suggests that the MTB in **D1** contains multiple quantum dots in series.

The above measurements were performed at $V_G = 0.42$ V, where the blockade effect is less significant [red dot in Fig. 2(d)]. We performed similar analysis at different gate voltages. We find that at gate voltages where the blockade is relatively weak, the power law exponents are between 0.7 ~ 0.9 with no obvious gate dependence as long as the temperature is above the Coulomb blockade-Luttinger liquid crossover. At gate voltages where the blockade is strong, we do not observe clear Luttinger liquid behavior at our measured temperatures. We discuss examples from these two regimes in Supplemental Material Fig. S2 [$V_G = -0.87$ V, pink dot in Fig. 2(d)] and S3 [$V_G = -1.56$ V, green dot in Fig. 2(d)] [13], respectively.

For the Type II MTB device **D2**, the zero-bias conductance at $V_G = 0$ V exhibits a single power law from 5 to 297 K [Fig. 3(b)], with an exponent of 2.1. The scaled differential conductance-scaled bias plot ($dI/dV/T^\alpha$ vs. $eV/k_BT$) at different temperatures and biases [Fig. 3(d)] again collapse into a single universal curve. The agreement between the temperature exponent and the bias exponent (2.0) provides strong evidence that the electrons in the Type II MTB form a Luttinger liquid with the parameter $g = 0.101$ (0.097), given that $\alpha = 2.0$ (2.1). The larger power-law exponent compared to that of **D1** indicates a stronger electron-electron interaction effect caused by the more frequent scattering slowing down the Fermi velocity (see Discussion below).

We also investigated the Luttinger liquid behaviors at different gate voltages for **D2**. Supplemental Material Figure S4 [13] shows that the device conductance follows a power-law behavior as a function of $T$, and the exponent increases as $V_G$ is more negative. At more positive $V_G$, the device behavior deviates from the power law: this is because at high doping the device conduction is dominated by the bulk crystal and thus the Luttinger liquid model fails. Supplemental Material Figure S5 [13] shows the scaled conductance plots at $V_G = -3$ V and $V_G = 8$ V.

**Discussion**

The transport behaviors of the two MTB types are different. For Type I, the power-law exponent is around 0.9 and weakly gate-dependent, persisting even when the Fermi level is deep in the bulk gap. In contrast, for the Type II MTB, the exponent is 2.0 and increases as the Fermi level goes into the bulk gap. These observations point to the significant role of atomic-scale defects in the transport behaviors.

The metallic nature of the MTB originates from its nontrivial topology. Because of the inversion symmetry breaking in $MoS_2$ monolayer, the K and K' valleys exhibit opposite valley Chern numbers of $\pm 1/2$ [15,31]. Therefore, an MTB produces a valley Chern number change of $\pm 1$ across the boundary, producing one pair of valley-projected counter-propagating states along the MTB [32]. Using the first-principles calculations (see Supplemental Material [13] for details), we obtain the relaxed $MoS_2$ bicrystal structure with one MTB [Fig. 4(a)] and its electronic band structure [Fig. 4(b)]. As shown in Fig. 4(b), such a structure features in-gap states. The calculated partial densities of the in-gap states reveal their association with the MTB, as illustrated in Fig. 4(a). The MTB band shown in Fig. 4(b) has opposite velocities at the projected K and K' valleys, indicating valley-chirality locking.

The intervalley scattering from atomic-scale scatterers prevents the valley from being a good quantum number and can localize the 1D MTB states [33]. This leads to a decrease in Fermi velocity or the degradation of the 1D states. As illustrated phenomenologically in Fig. 4(c), the increase of atomic-scale scattering causes the MTB band to be more localized (less dispersive), eventually producing an energy gap near the bulk valence band. For the Type I MTB with few scatterers [Fig. 4(c) left], the Fermi velocity (slope of the dispersion) is insensitive to the Fermi level position, and thus the power-law exponent, a measure of the electron interaction strength, shows minimal gate dependence. For the type II MTB with more scatterers arising from the frequent 60° turns [Fig. 4(c) right], the Fermi velocity continuously decreases as the Fermi level is gate tuned to deep in the bulk gap. The lower velocity enhances the electron interaction effect, which accounts for the observed larger power-law exponent as the gate becomes more negative.

The role of scatterers in MTB is reminiscent of a previous study of carbon nanotubes [34]. The transport properties of metallic tubes are not very sensitive to long-range scatterers because of the large-momentum separation between the two valleys. Nevertheless, atomic-scale scatterers can significantly modify the transport properties. One critical difference between the carbon nanotubes and MTBs is that both chiralites are present at each valley of the metallic carbon nanotubes, while the valley-chirality locking causes MTBs to be immune to intra-valley backscattering.

The 1D topological states similar to our MTB states were previously studied in bilayer graphene with a layer-stacking or electric-field domain wall [33,35-38]. In that case, however, an external electric field is required to break the inversion symmetry and produce a valley Chern number. Additionally, the electric field-induced gap in bilayer graphene is small, and the 1D electronic interaction effects such as Luttinger liquid behavior [4,5] and Coulomb blockade have not been observed, even at low temperature. In our $MoS_2$ bicrystal, the gap is intrinsic with a magnitude of ~2 eV, permitting the Luttinger liquid behavior on MTB to persist up to room temperature.

**Summary and outlook**

Our transport measurements on MTBs in monolayer $MoS_2$ bicrystals display Coulomb blockade and Luttinger liquid physics arising from strong electron-electron interactions. These measurements are enabled by the engineered MOCVD epitaxial growth, which yields MTBs with lengths of tens of micrometers, necessary for fabricating transistor-like devices for electrical characterizations. Our TEM characterizations and theoretical considerations shed light on the topological nature of the MTBs and reveal the importance of atomic-scale defects in electron transport along the 1D MTBs.

This work demonstrates that MTB hosts various novel physical phenomena, including single-electron transport, correlated electron states, and nontrivial band topology. These properties may be utilized in building novel electronic devices. By engineering the electrical contacts, the nontrivial topology may lead to dissipationless transport at room temperature because of the large bandgap in $MoS_2$. While this work features electrons tunneling into the 1D MTB channel due to large contact resistance, future improvements in contacts will enable discovery of more exotic physics in MTBs, such as the spin Kondo effect [39] and the valley-charge separation [4,5]. Finally, our growth method can be easily scaled up, providing a pathway to produce MTB circuits.

**Acknowledgement**

We acknowledge support from AFOSR (FA9550-21-1-0216), the DoD Vannevar Bush Faculty Fellowship (N00014-16-1-2825), NSF CUA (PHY-1125846), Samsung Electronics, NSF (PHY-1506284 for H.P., DGE-1745303 for E.B.), AFOSR MURI (FA9550-17-1-0002), ARL (W911NF1520067), and DOE (DE-SC0020115). M.-H.J. acknowledges support by the Institute for Basic Science (IBS), Korea, under Project Code IBS-R034-D1. The theoretical work done at UT Dallas was supported by NSF under Grants no. DMR-1945351, no. DMR-2105139, and no. DMR-2324033. We acknowledge the Texas Advanced Computing Center (TACC) for providing resources that have contributed to the research results reported in this work. K.W. and T.T. acknowledge support from the JSPS KAKENHI (Grant Numbers 20H00354 and 23H02052) and World Premier International Research Center Initiative (WPI), MEXT, Japan.

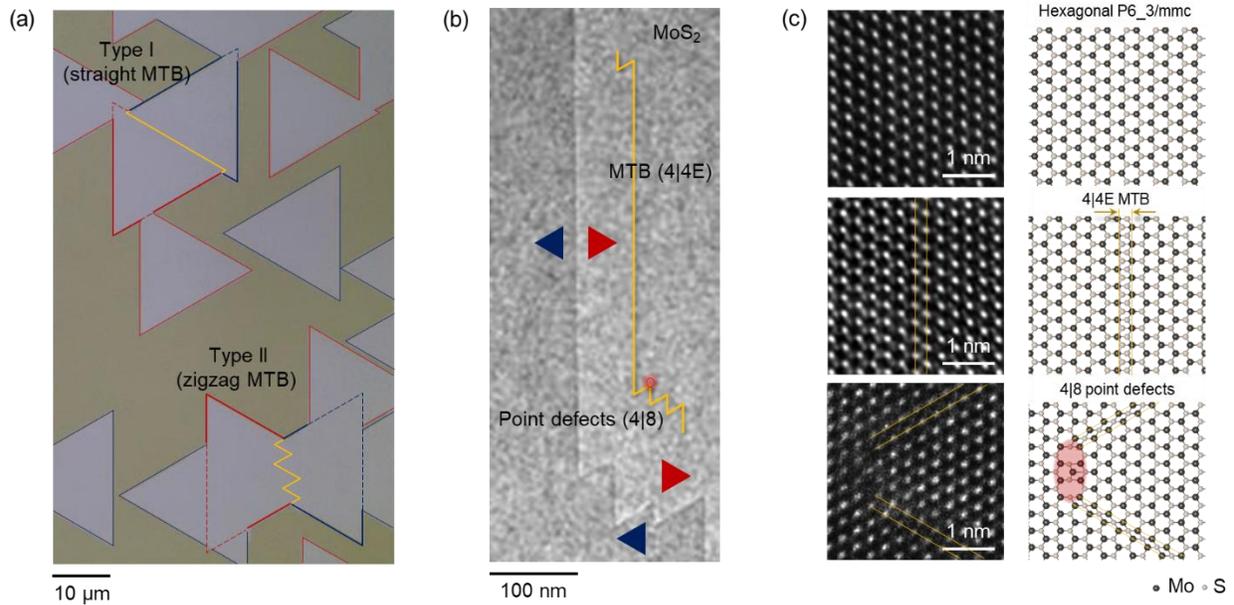

**FIG. 1. Epitaxial *vdW* MoS$_2$ monolayer bicrystals with three types of sublattices.** (a) Optical image of MoS$_2$ monolayer bicrystals grown on c-plane sapphire. Blue and red lines indicate two variants of MoS$_2$ monolayer and orange lines depict the two types of MTB. (b) Representative DF-TEM image obtained from a stitched MoS$_2$ monolayer bicrystal. The two variants (blue and red triangles) with different contrast in DF-TEM are stitched and form MTB. The orange line is a schematic of MTB. (c) In-plane HAADF-STEM image (left) obtained from MoS$_2$ monolayer bicrystals and the corresponding atomic models (right) for three types of sublattices: hexagonal MoS$_2$ (top), 4|4E MTB (middle), and point defects (bottom). Orange lines indicate typical MTBs.

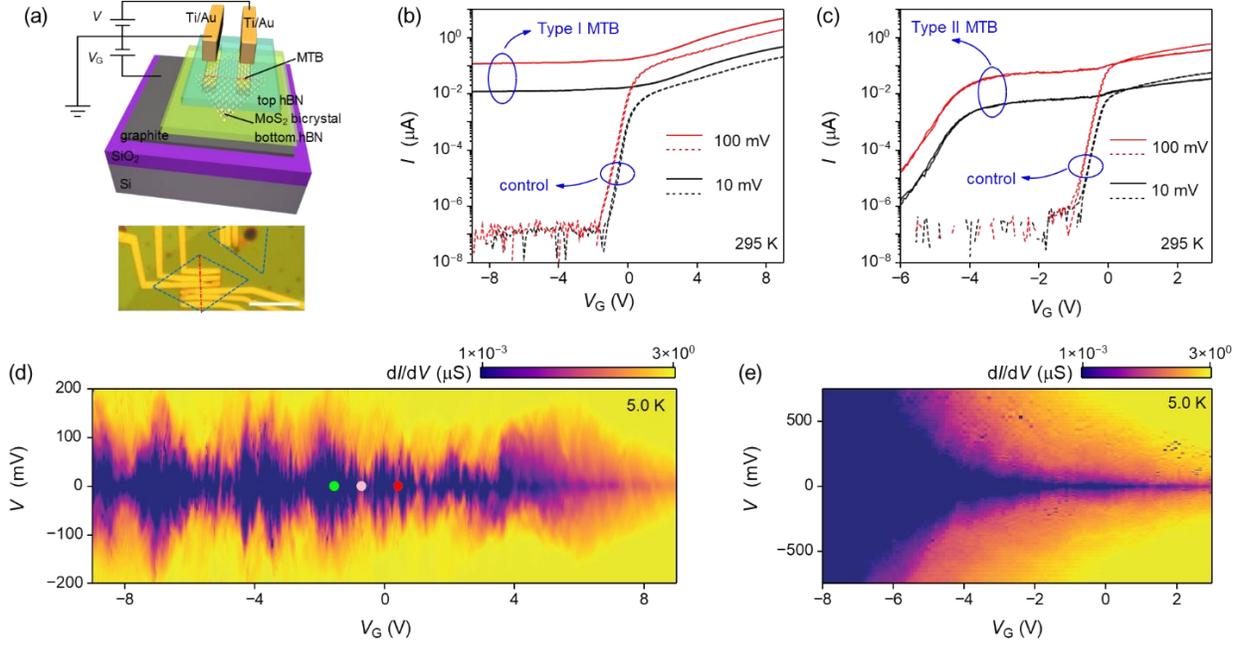

**FIG. 2. Transport properties of MTB devices.** (a) Top: a schematic showing the MTB device structure. Bottom: optical image of the Type I MTB devices and the control devices, where monolayer $MoS_2$ flakes are outlined by the blue dashed lines and the MTB by the red dash-dotted line. Scale bar: 10 µm. (b),(c) Transistor drain-source current $I$ as a function of $V_G$ for **D1** (b) and **D2** (c), together with the control devices, at room temperature. $V_G$ sweeps from negative maxima to positive maxima and then back. The legends show the applied biases $V$. (d),(e) $dI/dV$-$V$-$V_G$ maps for **D1** (d) and **D2** (e) at 5.0 K. **D1** displays aperiodic Coulomb blockade pattern. In (d), red ($V_G$ = 0.42 V) and pink dots ($V_G$ = -0.87 V) denote gate voltages where the blockade is less significant, while the green dot ($V_G$ = -1.56 V) denotes a gate voltage with strong blockade.

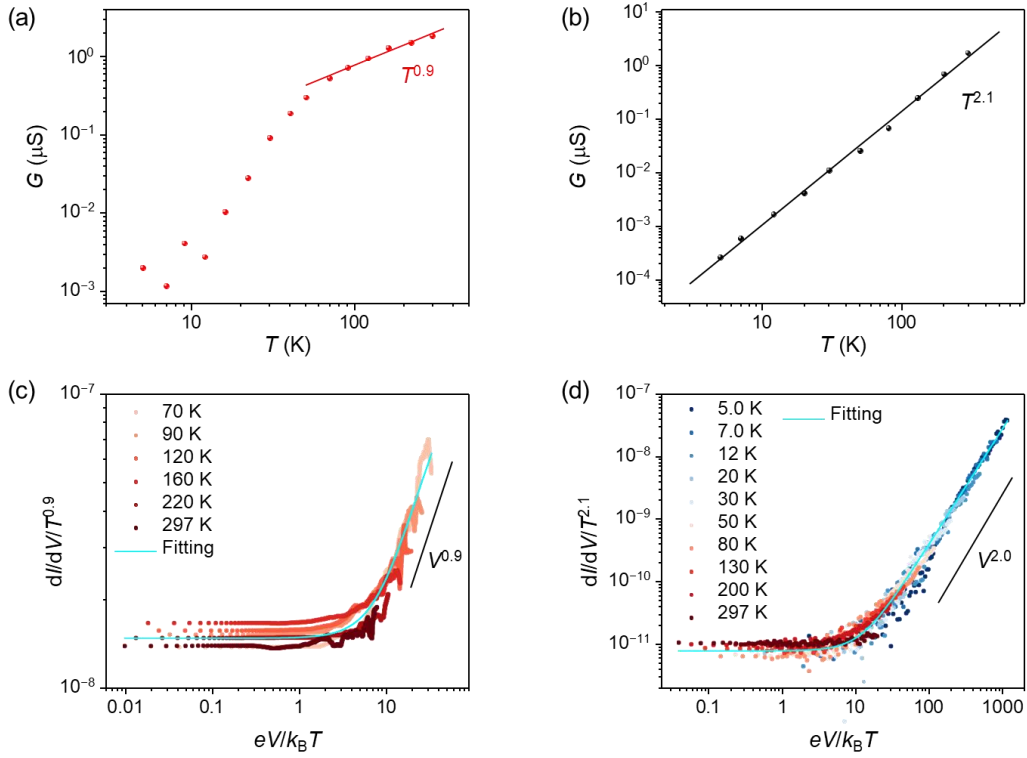

**FIG. 3. Luttinger liquid behaviors in MTBs.** (a),(b) Zero-bias conductance $G$ versus temperature $T$ of **D1** (a) and **D2** (b). For **D1**, above a crossover temperature, $G$ follows a $T^{0.9}$ power law. $G$ is rapidly suppressed below the crossover temperature. For **D2**, a single $T^{2.1}$ power law is observed. (c),(d) The scaled conductance versus the scaled bias plots of **D1** (c) and **D2** (d) for temperatures in the power law regime. All data collapse to single curves, showing $V^{0.9}$ and $V^{2.0}$ power laws at high biases for **D1** and **D2**, respectively.

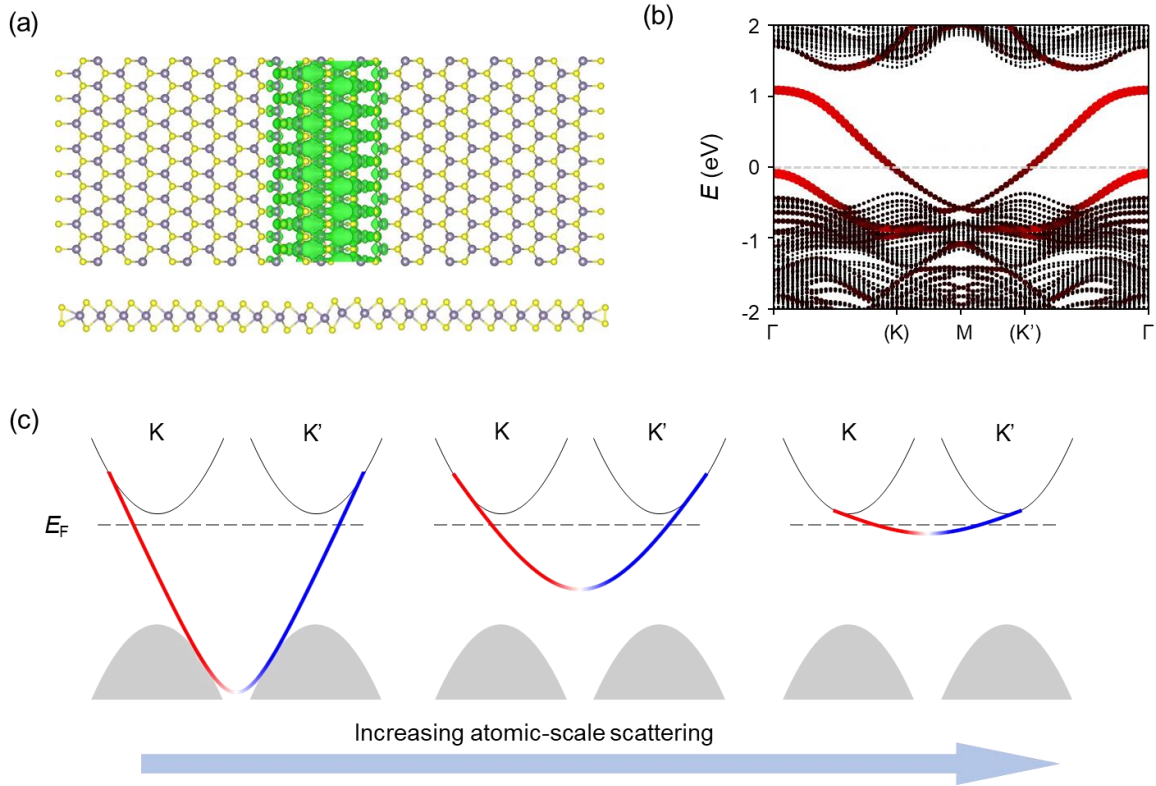

**FIG. 4. Topological origin of the metallic MTB and the role of atomic-scale scatterers.** (a) Top and side views of the calculated relaxed crystal structure of a $MoS_2$ bicrystal with one MTB. The in-gap states are associated with the MTB (green). (b) The calculated electronic band structure of the $MoS_2$ bicrystal with one MTB in (a), featuring valley-chirality locked 1D band in the bulk gap (red dots). (c) Schematics of the effect of atomic-scale scattering on altering the MTB band. The MTB band of the left and right chiralities are represented by the red and blue curves around the K and K' valleys, respectively.